\documentclass[preprint]{revtex4}
\usepackage{enumerate}
\usepackage{graphicx}
\usepackage{dcolumn}
\usepackage{bm}
\usepackage{amsmath}
\usepackage{amsxtra}
\usepackage{amstext}
\usepackage{amssymb}
\usepackage{latexsym}

\begin{document}

\title{
{Relaxation of Collisional Magnetically Confined Plasmas to Mechanical Equilibria}
}

\author{ Giorgio SONNINO}
\email{gsonnino@ulb.ac.be}
\affiliation{
Universit{\'e} Libre de Bruxelles (U.L.B.)\\
Department of Theoretical Physics and Mathematics\\
Campus de la Plaine C.P. 231 - Bvd du Triomphe\\
B-1050 Brussels - Belgium\\
}


\begin{abstract}
The relaxation of magnetically confined plasmas in a toroidal geometry is analyzed. From the equations for the Hermitian moments, we show how the system relaxes towards the mechanical equilibrium. In the space of the parallel generalized frictions, after fast transients, the evolution of  collisional magnetically confined plasmas is such that the projections of the evolution equations for the parallel generalized frictions and the shortest path on the Hermitian moments coincide. 
\noindent For spatially-extended systems, a similar result is valid for the evolution of the {\it thermodynamic mode} (i.e., the mode with wave-number ${\bf k} = {\bf 0}$).
\noindent The expression for the affine connection of the space covered by the generalized frictions, close to mechanical equilibria, is also obtained. The knowledge of the components of the affine connection is a fundamental prerequisite for the construction of the (nonlinear) closure theory on transport processes.

\end{abstract}

\maketitle


\section{Introduction}
\label{intro}
It is known that a macroscopic description of thermodynamic systems requires the formulation of a theory for the {\it closure relations}. A thermodynamical field theory (TFT) has been established in order to determine the (non linear) deviations from of the Onsager coefficients for thermodynamic systems far from equilibrium \cite{sonninog}. The nonlinear transport equations have been derived by imposing that the thermodynamic theorems for systems out of equilibrium \cite{prigogine1} and the De Donder-Prigogine principle, also refereed to as the {\it Thermodynamic Covariance Principle} (TCP) \cite{dedonder} (see the footnote \footnote{The TCP establishes that: "Thermodynamic systems are thermodynamically equivalent if, under transformation of fluxes and forces, the bilinear form of the entropy production remains unaltered". Flux-forces transformations leaving invariant the expression of the entropy production, are referred to as the {\it Thermodynamic Coordinate Transformations} (TCT).}), are respected.

\noindent Magnetically confined tokamak plasmas are a typical example of thermodynamic systems out of Onsager's region. In this case, even in absence of turbulence, the local distribution functions of species (electrons and ions), as well as the distribution function for the fluctuations of the thermodynamic quantities, deviate from the (local) Maxwellian. According to the Onsager theory of fluctuations, in this condition, the thermodynamic fluxes are not (in general) linearly connected with the conjugate forces (ref. to the Onsager theory [4] and, for example, [5]. See also the end of section \ref{cs}). In tokamak plasmas, the thermodynamic forces and the conjugate flows are the generalized frictions and the Hermitian moments, respectively \cite{balescu2}. The neoclassical theory is a {\it linear transport theory}  (see, for example, \cite{balescu2}) meaning by this, a theory where the moment equations are coupled to the closure relations (i.e. flux-force relations), which have been {\it linearized with respect to the generalized frictions} (see, for example, Ref. \cite{balescu1}). This approximation is in contrast with the fact that the distribution function of the thermodynamic fluctuations is not a Maxwellian and could provoke some disagreements with the experimental profiles \cite{sonnino3}, \cite{sonnino4}. It is, however, important to mention that it is well accepted that the main reason of this discrepancy is attributed to turbulent phenomena existing in tokamak plasmas. Fluctuations in plasmas can become unstable and therefore amplified, with their nonlinear interaction, successively leading the plasma to a state, which is far away from equilibrium. In this condition, the transport properties are supposed to change significantly and to exhibit qualitative features and properties that could not be explained by collisional transport processes, e.g. size-scaling with machine dimensions and non-local behaviors that clearly point at turbulence spreading etc. (see, for example, Ref.~\cite{diamont}).

\noindent The nonlinear transport equations have been largely used for studying transport processes in non equilibrium systems such as  magnetically confined plasmas, materials submitted to temperature and electric potential gradients or chemical reactions. In magnetically confined plasmas, the nonlinear transport equations provide a link between the generalized frictions (the thermodynamic forces) and the Hermitian moments (the conjugate flows). This allows to determine the particle fluxes (electrons and ions) and energy losses as well as the (nonlinear) particle distribution functions. This task has been accomplished in Ref. \cite{sonnino3} and in the papers reported in Ref.~\cite{sonnino4} and in the footnote \footnote{Sonnino G. and Peeters P., {\it Nonlinear Transport Processes in Tokamak Plasmas. Part II: The low-collisional regimes}, submitted to publication in the review {\it Physics of Plasmas} (2011).}.

\noindent Finally, the nonlinear transport equations have been derived by adding a further assumption: {\it There exists a thermodynamic action, scalar under TCT, which is stationary for general variations in the transport coefficients and the affine connection of the thermodynamical forces space} \cite{sonninog}. However, the determination of this action requires the knowledge of the affine connection. The expression of the affine connection can be derived by analyzing several examples of relaxation. 

\noindent In the manuscript reported in Ref.~\cite{sonninojmp}, we analyze the relaxation of chemical reactions to stable steady-states in the chemical affinities space. In these cases, the small parameter $\epsilon$ measures the distance of the system from the steady state. 

\noindent In this paper, we shall be concerned with the relaxation of magnetically confined plasmas in a toroidal geometry. The characteristic feature of the evolution equations is the presence of a small parameter $\epsilon$, the {\it drift parameter}, defined as the Larmor radius over a macroscopic length \cite{balescu2}.  In this situation, the long-time behaviour of the solution, describing the evolution of the system near the steady state, may be obtained by using the multiple time-scale perturbation expansion (see, for example, the book cited in Ref. \cite{davidson}). Starting from the balance equations for mass, energy and higher order hermitian moments, applied to magnetically confined plasmas, we show, in Section \ref{plasma}, the validity of the following theorems: 
\begin{description}
\item {($a$) After fast transients, in the Onsager region of the generalized frictions space, a homogeneous system relaxes towards the mechanical equilibrium along a straight line.} 
\item {($b$) After fast transients, out of the Onsager region, a homogeneous system relaxes towards a stable mechanical equilibrium such that $J_\mu\ \mathcal{U}^\mu (X_{ev.-tr.},\varrho) - J_\mu\ {\mathcal U}^\mu(X_{s.path},\varrho)=O(\epsilon^2)$ where $\mathcal{U}^\mu (X_{ev.-tr.},\varrho)$ and ${\mathcal U}^\mu(X_{s.path},\varrho)$ are the evolution equation for the generalized frictions and the shortest path equation in the generalized frictions space, respectively. $X^\mu$ and $J_\mu$ denote the vector of the generalized frictions and the vector of the Hermitian moments, respectively. The trajectory traced out by the system and the shortest past are parametrized by $\varrho$. This parameter is defined in the Subsection \ref{plasma1}}
\end{description}
\noindent We assume that the mechanical equilibrium is a steady-state of the plasma. Similar theorems for the relaxation of the {\it thermodynamic mode} (i.e., the mode with wave-number ${\bf k} = {\bf 0}$) to the mechanical equilibrium, can also be derived for spatially-extended plasmas.

\noindent Even though, the example examined in this paper refers to the relaxation of magnetically confined plasmas in a toroidal geometry, the results obtained are valid generally because the dynamics include all relevant moment equations and the parameter $\epsilon$ is not related to the distance of the system from the stationary states.  Such example enable us to determine the expression of the affine connection for the thermodynamical forces space, near the non-equilibrium steady-states. This task is accomplished in Section \ref{cs}.
\vskip 0.2truecm
\section{Relaxation of Magnetically Confined Plasmas}\label{plasma}
In this section we analyze the case of magnetically confined plasmas. For simplicity, we consider {\it fully ionized plasmas} defined as a collection of electrons and positively charged ions \cite{balescu1}. The balance moment equations for {\it mass}, {\it heat} and the {\it non-priviledged} fifth hermitian moments read, respectively \cite{balescu2}
\begin{align}\label{c1}
\!\!\!\!\!\!\!\!\!\!\!\!\!\!\!\!\!\!\!\!
 \tau_\alpha\partial_t q_r^{\alpha(1)}=&\ \Omega_\alpha \tau_\alpha\epsilon_{rmn}
 q^{\alpha(1)}_mb_n+\tau_\alpha Q_r^{\alpha(1)}
+g_r^{\alpha(1)}+{\bar g}_r^{\alpha(1)}+O(\epsilon^2)\nonumber\\
\!\!\!\!\!\!\!\!\!\!\!\!\!\!\!\!\!\!\!\!
 \tau_\alpha\partial_t q_r^{\alpha(3)}=&\ \Omega_\alpha \tau_\alpha\epsilon_{rmn}
 q^{\alpha(3)}_mb_n+\tau_\alpha Q_r^{\alpha(3)}
 +g_r^{\alpha(3)}+{\bar g}_r^{\alpha(3)}+O(\epsilon^2)
 \\
\!\!\!\!\!\!\!\!\!\!\!\!\!\!\!\!\!\!\!\!
 \tau_\alpha\partial_t q_r^{\alpha(5)}=& \Omega_\alpha \tau_\alpha\epsilon_{rmn}+q^{\alpha(5)}_mb_n+\tau_\alpha Q_r^{\alpha(5)}
 +{\bar g}_r^{\alpha(5)}+O(\epsilon^2)\nonumber
 \end{align}

\noindent where $q_r^{\alpha (n)}$ denote the {\it Hermitian moments of the distribution functions}, $Q_r^{\alpha(1)}$ indicate the {\it dimensionless generalized collisional friction terms} and $g_r^{\alpha (n)},\ {\bar g}_r^{\alpha (n)}$ are the {\it dimensionless source terms}. The suffix $\alpha$ distinguishes the two electric species i.e., $\alpha=e$ for electrons and $\alpha=i$ for ions. Index $n$ takes the values $n=(1,3,5)$. Moreover, $\Omega_{\alpha}$ and $\tau_{\alpha}$ are the {\it Larmor frequency} and the {\it relaxation time} of species $\alpha$, respectively. $\epsilon_{rmn}$ and $b_r$ denote, respectively, the {\it Levi-Civita tensor} and the {\it unit magnetic field} components. In this paper, the summation convention over repeated indices is understood. On the contrary, in Eqs~(\ref{c1}), as well as in what follows, there is {\it no} summation over the index $\alpha$.  Eqs~(\ref{c1}) are the moment equations describing collisional plasmas in the presence of {\it an inhomogeneous and curve magnetic field}. These plasmas are referred to as (magnetically confined) {\it tokamak plasmas}. In this paper we deal with {\it collisional transport processes} (i.e., not turbulent plasmas) where collisions are the only source of irreversibility or dissipation. In this case, we can distinguish two collisional transport regimes: the fully collisional transport regime (or the Pfirsch-Schl{\"u}ter transport regime) and the low-collisional transport regimes (or the plateau and the banana regimes). Physically, these two transport regimes are distinguished by the order of magnitude of $\lambda_{mfp}$, the mean free path. When the mean free path is much smaller than the macroscopic length scale $L_M$ we are in the so-called short mean free path regime or, the {\it fully collisional transport regime} (fctr). The macroscopic length is associated with the gradient lengths of the macroscopic quantities, such as the density gradient, the temperature gradient, etc.. $L_M$ is defined as the shortest of these lengths. On the other hand, if the fusion temperatures are realized, the mean free path $\lambda_{mfp}$ can easily be made to exceed $L_M$. In this case we are in the so-called long mean free path regime or the {\it low-collisional transport regime} (lctr). In this regime, the orbits of the particles are bounded in the space (the {\it banana} orbits) (for easy references, see for example Refs \cite{balescu2}). A useful transport theory must cover both the fully collisional and the low-collisional transport regimes. By a {\it magnetized plasma} it is meant a plasma where the scale length characterizing the plasma (in general the hydrodynamic scale) is much larger than the gyro-radii of its constituent charged particles. Thus, by defining with $\rho_L$ the {\it Larmor radius} (the radius of the circular motion of a charged particles in the presence of a uniform magnetic field), the plasma is magnetized if the parameter
\begin{equation}\label{c3}
\epsilon\equiv\frac{\rho_L}{L_M}
\end{equation}
\noindent is much less than one. Wherever is valid the approximation $\epsilon\ll 1$, we are within the so-called {\it drift parameter approximation}. It can be shown that plasmas in the both transport collisional regimes, described above, can be validly treated in the drift approximation [see, for example, ref. \cite{balescu2}]. The presence of the small parameter $\epsilon$ allows to neglect in the equations all terms of order $\epsilon^2$ or higher. 
\begin{figure}\resizebox{0.45\textwidth}{!}{%
\includegraphics{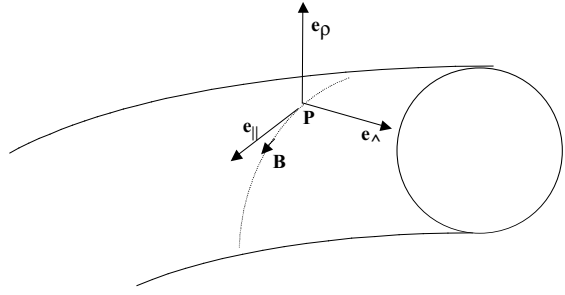}}
\caption{ \label{localtriad} The local dynamical triad. The {\it radial direction} ${\bf e_{\rho}}$ is the unit vector radially directed pointing outwards. The {\it parallel direction} ${\bf e_{\parallel}}$ coincides with the unit magnetic vector i.e., ${\bf e_{\parallel}}\equiv {\bf B}/B$. The right-handed coordinate system is obtained defining the {\it perp-tangential direction} ${\bf e_{\wedge}}$ as ${\bf e_{\wedge}}={\bf e_{\rho}} \wedge{\bf e_{\parallel}}$.}
\end{figure}
\noindent In Ref.~\cite{balescu2} an intrinsic kinetic form of the dimensionless (density of) entropy production $\sigma^\alpha$ of species $\alpha$ is derived under the sole assumption that the state of the plasma is not too far from the reference local equilibrium state. It is shown that, in the {\it local dynamical triad} [see Fig.~(\ref{localtriad})], the entropy production is closely associated with the collision term and it can be brought into the form
\begin{align}\label{c2}
\sigma^e=&\ q_{\parallel ps}^{(1)}(g_{\parallel}^{(1)}-{\bar g}_{\parallel}^{e(1)})+
q_{\parallel ps}^{e(3)}(g_{\parallel}^{e(3)}+{\bar g}_{\parallel}^{e(3)})
+q_{\parallel bpcl}^{(1)}(g_{\parallel}^{(1)}-{\bar g}_{\parallel}^{e(1)})
+q_{\parallel bpcl}^{e(3)}(g_{\parallel}^{e(3)}+{\bar g}_{\parallel}^{e(3)})\nonumber\\
&+q_{\parallel bpcl}^{e(5)}{\bar g}_{\parallel}^{e(5)}+{\hat q}_{\rho cl}^{e(1)}g_{\rho}^{(1)P}+
{\hat q}_{\rho cl}^{e(3)}g_{\rho}^{e(3)}\\
\sigma^i=&\ q_{\parallel ps}^{i(3)}(g_{\parallel}^{i(3)}+{\bar g}_{\parallel}^{i(3)})\ +q_{\parallel bpcl}^{i(3)}(g_{\parallel}^{i(3)}+{\bar g}_{\parallel}^{i(3)})
+q_{\parallel bpcl}^{i(5)}{\bar g}_{\parallel}^{i(5)}+{\hat q}_{\rho cl}^{i(3)}g_{\rho}^{i(3)}\nonumber
\end{align}
\noindent For easy reference, we list here explicitly the relations between the dimensionless and the corresponding dimensional Hermitian moments:
\begin{eqnarray}\label{entropy1}
&&q_r^{\alpha (1)}=\Bigl(\frac{m_{\alpha}}{T_{\alpha}}\Bigr)^{1/2}\frac{1}{n_{\alpha}}\Gamma_r^{\alpha}\nonumber\\
&&q_r^{(1)}=\frac{1}{en_e}\Bigl(\frac{m_e}{T_e}\Bigr)^{1/2}j_r\nonumber\\
&&q_r^{\alpha (3)}=\sqrt{\frac{2}{5}}\Bigl(\frac{m_{\alpha}}{T_{\alpha}}\Bigr)^{1/2}\frac{1}{T_{\alpha}n_{\alpha}}Q_r^{\alpha}\\
&&q_r^{\alpha (5)}=\frac{1}{n_{\alpha}}\Bigl(\frac{m_{\alpha}}{T_{\alpha}}\Bigr)^{1/2}L_r^{\alpha}
\qquad\quad\qquad{\rm with}\qquad r=(\rho,\parallel,\wedge)
\nonumber
\end{eqnarray}
\noindent where $m_{\alpha}$, $n_\alpha$ and $T_{\alpha}$ are the mass, the number density and the temperature of species $\alpha$, respectively. Moreover, $j_r$, $\Gamma_r^{\alpha}$ and $Q_r^{\alpha}$ indicate the {\it electric current}, the {\it particle fluxes} and the {\it heat fluxes}, respectively. $L_r^{\alpha}$ is the dimensional fifth-order Hermitian moment corresponding to $q_r^{\alpha (5)}$. For completeness, we also report the relation between the {\it pressure tensor} $\pi_{rs}^{\alpha}$ and the {\it second-order tensor Hermitian moment} $q_{rs}^{\alpha (2)}$
\begin{equation}\label{entropy2}
q_{rs}^{\alpha (2)}=\frac{1}{\sqrt{2}n_{\alpha}T_{\alpha}}\pi_{rs}^{\alpha}
\qquad\quad\qquad{\rm with}\quad r,s=(\rho,\parallel,\wedge)
\end{equation}
\noindent The dimensionless source terms are defined as
\begin{eqnarray}\label{entropy3}
&&\!\!\!\!\!\!\!\!\!\!\!\!\!g_r^{(1)}=\tau_e\Bigl(\frac{m_e}{T_e}\Bigr)^{1/2}\Bigl(\frac{e}{m_e}E_r-\Omega_e\epsilon_{rmn}u_mb_n+\frac{1}{m_en_e}\nabla_r(n_eT_e)\Bigr)\nonumber\\
&&\!\!\!\!\!\!\!\!\!\!\!\!\!g_r^{\alpha (1)}=\tau_{\alpha}\Bigl(\frac{m_{\alpha}}{T_{\alpha}}\Bigr)^{1/2}\Bigl(\frac{e_{\alpha}}{m_{\alpha}}E_r-\frac{1}{m_{\alpha}n_{\alpha}}\nabla_r(n_{\alpha} T_{\alpha})\Bigr)\nonumber\\
&&\!\!\!\!\!\!\!\!\!\!\!\!\!g_r^{\alpha (3)}=-\sqrt{\frac{5}{2}}\tau_{\alpha}\Bigl(\frac{T_{\alpha}}{m_{\alpha}}\Bigr)^{1/2}\frac{1}{T_{\alpha}}\nabla_rT_{\alpha}\nonumber\\
&&\!\!\!\!\!\!\!\!\!\!\!\!\!{\bar g}_r^{\alpha (1)}=-\sqrt{2}\tau_{\alpha}\Bigl(\frac{m_{\alpha}}{T_{\alpha}}\Bigr)^{1/2}\frac{1}{m_{\alpha}n_{\alpha}}\nabla_s(n_{\alpha}T_{\alpha}q_{rs}^{\alpha (2)})\\
&&\!\!\!\!\!\!\!\!\!\!\!\!\!{\bar g}_r^{\alpha (3)}=-\sqrt{\frac{2}{5}}\tau_{\alpha}\Bigl(\frac{T_{\alpha}}{m_{\alpha}}\Bigr)^{1/2}\!
\Bigl[\sqrt{7}\frac{1}{n_{\alpha}T^2_{\alpha}}\nabla_s(n_{\alpha}T^2_{\alpha}q_{rs}^{\alpha (4)})\!+\!\sqrt{2}T_{\alpha}^{-7/2}\nabla_s(T_{\alpha}^{7/2}q_{rs}^{\alpha (2)})\Bigr]\nonumber\\
&&\!\!\!\!\!\!\!\!\!\!\!\!\!{\bar g}_r^{\alpha (5)}=-\sqrt{\frac{2}{5}}\tau_{\alpha}\Bigl(\frac{T_{\alpha}}{m_{\alpha}}\Bigr)^{1/2}\Bigl[
\frac{3}{n_{\alpha}T^3_{\alpha}}\nabla_s(n_{\alpha}T^3_{\alpha}q_{rs}^{\alpha (6)})\nonumber\\
&&\qquad\qquad\qquad\qquad\qquad+2T_{\alpha}^{-11/2}\nabla_s(T_{\alpha}^{11/2}q_{rs}^{\alpha (4)})+
\sqrt{14}q_{rs}^{\alpha (2)}T_{\alpha}^{-1}\nabla_sT_{\alpha}\Bigr]\nonumber
\end{eqnarray}
\noindent with $r=(\rho,\parallel,\wedge)$ and $q_{rs}^{\alpha (2n)}$ are the traceless tensorial Hermitian moments. Indicating with $e$ the {\it absolute value of the charge of the electron} and with $Z$ the {\it charge number} of the ions, we have $e_{\alpha}=-e$ for electrons and $e_{\alpha}=+Ze$ for ions. {\bf E} and ${\bf u}$ are the electric field and the centre-of-mass velocity of the plasma, respectively. Moreover $b_n={\bf B}/B$ and $\epsilon_{rmn}$ is the Levi-Civita symbol. Eqs~(\ref{c2}) allows us to identify the {\it thermodynamic forces} $X^{\alpha\mu}$ and the {\it thermodynamic flows} $J^\alpha_\mu$:
\begin{equation}\label{c2a}
X^{\alpha\mu}\equiv g_{\parallel}^{\alpha(n)}+{\bar g}_{\parallel}^{\alpha(n)}\ ;\ 
J_\mu^\alpha\equiv  q_{\parallel}^{\alpha(n)}\  \mu= (1,2,3),\ n=(1,3,5)
\end{equation}
\noindent Hence, the thermodynamic forces are related to the spatial inhomogeneity and they are expressed as gradients of the thermodynamic quantities whereas the thermodynamic flows are the parallel component of the quantities defined in Eq.~(\ref{entropy1}). Different, but also very important, quantities are the {\it transport fluxes} defined as the radial fluxes averaged over a magnetic surface. These fluxes are denoted by $<q_\rho^{\alpha(n)}>$, where $<\cdots>$ indicates the magnetic- surface averaging operation. In this notation, $<q_\rho^{\alpha(1)}>$ (i.e. $n=1$) and $<q_\rho^{\alpha(3)}>$ (i.e. $n=3$) denote the (dimensionless), magnetic averaged, radial particles flux and radial heat flux, respectively. Let us stress that the transport fluxes are different from the conjugate thermodynamic flows. This main conceptual step deserves some additional explanations. The thermodynamic forces, given defined in Eq.~(\ref{c2a}), are not directly accessible to measurement in a tokamak. Only surface-averaged quantities, depending solely on the radial coordinate $\rho$ are experimentally accessible. Typically, experiments provide us with temperature, density or pressure profiles and with transport fluxes. The characteristic feature of the neoclassical expression of the surface-averaged entropy production for toroidally confined {\it fully collisional} plasmas, is 
\begin{eqnarray}\label{c2a0}
&&<\sigma_e>_{fctr}=<q_\rho^{e(1)}>_{fctr}g_\rho^{(1)P}+<q_\rho^{e(3)}>_{fctr}g_\rho^{e(3)}
\nonumber\\
&&<\sigma_i>_{fctr}=<q_\rho^{i(3)}>_{fctr}g_\rho^{i(3)}
\end{eqnarray}
\noindent where $g_\rho^{(1)P}$ is the dimensionless radial pressure gradient (denoted by $\nabla_\rho P$)
\begin{equation}\label{c2a1}
g_\rho^{(1)P}=-\tau_e\Bigl(\frac{m_e}{T_e}\Bigr)^{1/2}\frac{1}{m_en_e}\nabla_\rho P
\end{equation}
\noindent Thus, the average entropy production, depending on the parallel components of the entering quantities, is expressed entirely in terms of the transport fluxes and radial forces. This inversion of roles is a direct consequence of the zero-divergence geometrical constraint. The latter transforms the average entropy production into a radial contribution. However, the thermodynamic form shown by Eqs~(\ref{c2a0}) is also due to the fact that, in the fully collisional regime, there is no contribution of the unprivileged moments, $q_{\parallel bpcl}^{\alpha(5)}$, to the average entropy production. This is not the case for the low-collisional transport regime. As a result, in this latter regime, the average entropy production does not derive from a thermodynamic form, which would have been expected \cite{balescu3}
\begin{eqnarray}\label{c2ab}
&&<\sigma_e>_{lctr}\ \neq\ <q_\rho^{e(1)}>_{lctr}g_\rho^{(1)P}+<q_\rho^{e(3)}>_{lctr}g_\rho^{e(3)}+{\hat q}_{\parallel lctr}^{(1)}\ {\hat g}_\parallel^{(1)A}
\nonumber\\
&&<\sigma_i>_{lctr}\ \neq\ <q_\rho^{i(3)}>_{lctr}g_\rho^{i(3)}
\end{eqnarray}
\noindent where ${\hat q}_{\parallel}^{(1)}$ and ${\hat g}_\parallel^{(1)A}$ are the {\it average dimensionless parallel electric current density} and the {\it average dimensionless induced external parallel electric field} (denoted with $E_\parallel^A$), respectively
\begin{eqnarray}\label{c2ac}
&&{\hat q}_\parallel^{(1)}=<\frac{B}{\beta_0}q_\parallel^{(1)}>\qquad ; \quad 
q_\parallel^{(1)}\equiv \Bigl(\frac{m_eT_i}{m_iT_e}\Bigr)^{1/2}q_\parallel^{i(1)}-q_\parallel^{e(1)}
\nonumber\\
&&{\hat g}_\parallel^{(1)A}=\Bigl(\frac{m_e}{T_e}\Bigr)^{1/2}\tau_e\frac{e}{m_e\beta_0}<BE_\parallel^A>\qquad ; \quad \beta_0=\sqrt{<B^2>}
\end{eqnarray}
\noindent In literature, this phenomenon is referred to as the {\it divorce between entropy production and transport theory for the banana transport mechanism}. Further physical and mathematical details concerning this subject can be found in the paper cited in the footnote \footnote{Sonnino G. and Peeters P., {\it Nonlinear Transport Processes in Tokamak Plasmas.Part II: The low-collisional Regimes}, submitted to publication in the review {\it Physics of Plasmas (PoP)} (2011)\label{PoP2}}. The transport fluxes are related to the parallel components of the dimensionless generalized frictions and the dimensionless source terms by [see, for example, Ref. \cite{balescu2})]
\begin{equation}\label{entropy4}
\left\{ \begin{array}{ll}
<{q}_{\rho}^{\alpha (n)}>_{fctr}=-K_{\alpha}\tau_{\alpha}<\frac{\beta_0}{B}\Bigl(1-\frac{B^2}{\beta_0^2}\Bigr)Q_{\parallel}^{\alpha (n)}>
&\quad\quad\mbox{in the fully collisional regime}\\
<{q}_{\rho}^{\alpha (n)}>_{lctr}=K_{\alpha}<\frac{B}{\beta_0}{\bar g}_{\parallel}^{\alpha (n)}>
&\quad\quad\mbox{in the low-collisional regime}
\end{array}
\right.
\end{equation}
\noindent In general, the quantity $K_{\alpha}$ is not a surface quantity but it becomes one for some configurations of the magnetic field as, for example, in the standard model. As previously mentioned, Eq.~(\ref{entropy4}) is a direct consequence of the zero-divergence constraint, introduced by the confining geometry, which couples the flux components in a new way and introduces the effect of the parallel thermodynamic forces on the transport fluxes. The evolution equations of the moments, Eqs~(\ref{c1}), have a hierarchical structure: the equations for the third moments will involve the fourth moments, and so. The dynamics of a thermodynamic system is finally based on the set of balance equations Eqs~(\ref{c1}) coupled to the {\it closure equations} \footnote{In this paper, we distinguish the {\it closure equations} from the {\it transport equations}. The evolution equations are {\it closed} by the {\it closure relations} (in our case, the relations between the parallel generalized collision friction terms $Q_\parallel^{\alpha(n)}$ and the flows $q_\parallel^{\alpha(n)}$) while the {\it transport equations} are the relations between the thermodynamic flows $J_\mu^\alpha (\equiv q_\parallel^{\alpha(n)}$) and the thermodynamic forces $X^{\alpha\mu}$ that produce them. \label{slaving}}, relating the dimensionless generalized collisional friction terms to the thermodynamic fluxes. In this example, the closure equations can be cast into the form \cite{balescu1} 
\begin{equation}\label{c2b}
\tau_{\alpha}Q_{\parallel}^{\alpha (n)}=-{\tilde g}^{mn}q_{\parallel}^{\alpha(m)}\qquad (m,n=1,3,5)
\end{equation}
\noindent where ${\tilde g}_{mn}$ (for simplicity, we omit the suffix $\alpha$) is a positive-definite matrix, which may depend on the thermodynamic forces $X^{\alpha\mu}$. In case of magnetically confined plasmas, this matrix is symmetric \cite{balescu2}. Inserting Eqs~(\ref{c2b}) into Eqs~(\ref{c1}) we obtain the balance equations for the thermodynamic forces
\begin{eqnarray}\label{c3}
&&\tau_\alpha\partial_tJ^\alpha_\mu=-\delta_{\mu\kappa}\bigl({\tilde g}^{\kappa\nu}J^\alpha_\nu-X^{\alpha\kappa}\bigr)
+O(\epsilon^2)
\ \  {\rm where}\\
&&
\delta_{\mu\kappa}=
\left\{ \begin{array}{ll}
0& \mbox{if $ \mu\neq \kappa$}\\
1& \mbox{if $ \mu= \kappa$}
\end{array}
\right.
\end{eqnarray}
\noindent The task of the transport theory is to describe the evolution of the gradients (of density, temperature, magnetic field etc.) towards global mechanical equilibrium. As long as the step length in the random walk is smaller than the gradient length scale this process is diffusive, so the associated time derivative is expected to be of the order 
\begin{equation}\label{c3a0}
\frac{\partial}{\partial t}\sim \frac{D}{L_M^2}\sim\frac{\epsilon^2}{\tau_\alpha}
\end{equation}
\noindent where $D\sim \rho_L^2/\tau_\alpha$ denotes the diffusion coefficients. Thus, for a magnetized plasma, very close to the steady-state, the partial time derivative in Eq.~(\ref{c3}) is assumed to be small in the sense \cite{helander}
\begin{equation}\label{c3a00}
\tau_\alpha\frac{\partial}{\partial t}\sim \epsilon^2
\end{equation}
\noindent The weaker version 
\begin{equation}\label{c3a000}
\tau_\alpha\frac{\partial}{\partial t}\sim \epsilon
\end{equation}
\noindent describes the relaxation of the plasma, evolving towards the steady-state on a shorter time-scale. To sum up our discussion, during the relaxation, close to the steady-state, we require that
\begin{description}
\item  {\bf i)} In the transport regimes (i.e., neglecting fast transients), the time derivatives $\tau_\alpha\partial_tJ^\alpha_\mu$ are assumed to be of order $\epsilon^2$. During the relaxation, near the steady-states, $\tau_\alpha\partial_tJ^\alpha_\mu$ are of order $\epsilon$. The weaker versions of the rate of the change in time represent the fast evolution of the system \cite{balescu2}.
\item {\bf ii)} The Universal Criterion of Evolution theorem (UCE), established by Glansdorff-Prigogine  \cite{prigogine1}, and the covariance properties of the evolution equations, with respect to the thermodynamic forces transformations \cite{sonninog}, are satisfied.
\end{description}
\noindent Finally, we recall that we are allowed to neglect in the equations all terms of order $\epsilon^2$ or higher.
\subsection{Homogeneous Plasmas}\label{plasma1}

\noindent In this case, Eqs~(\ref{c3}) may be rewritten, in a more convenient way, as 
\begin{equation}\label{c3a}
\begin{split}
\Bigl(\delta^\lambda_\mu+g^{\lambda\eta}X^{\alpha\kappa} g_{\mu\kappa,\eta}\Bigr)\tau_\alpha g_{\lambda\nu} {\dot X}^{\alpha\nu}=&-
\delta_{\mu\kappa}\bigl({\tilde g}^{\kappa\nu}J^\alpha_\nu-X^{\alpha\kappa}\bigr)
+O(\epsilon^2)
\end{split}
\end{equation}
\noindent where $\delta^\lambda_\mu$ denoting the Kronecker delta and we have introduced the {\it transport equations} relating the thermodynamic flows with the thermodynamic forces that produce them, through the {\it matrix of the transport coefficients}, $g_{\mu\nu}$ (we omit the suffix $\alpha$):
\begin{equation}\label{c3a0000}
J_\nu^\alpha= g_{\mu\nu}X^{\alpha\mu}
\end{equation}
\noindent In principle, the matrix of the transport coefficients may depend on $X^{\alpha\mu}$. Quite naturally, in solving this equation close to the steady-state, we should make an optimum use of the presence of the small parameters. In a magnetized plasma, in accord with experiment, the gyrofrequency $\Omega$ exceeds the collision frequency $\nu$,
\begin{equation}\label{c3a01}
\frac{\nu}{\Omega}\ll 1
\end{equation}
\noindent Taking into account that the mean-free path $\lambda=v_T\nu$ and $\rho_L=v_T/\Omega$, where $V_T$ indicates the thermal velocity, we obtain 
\begin{equation}\label{c3a02}
\Delta=\frac{\rho_L}{\lambda}
\end{equation}
\noindent So, Eq.~(\ref{c3a01}) again expresses the smallness of the Larmor radius.  A condition, which almost always pertains in practice, is
\begin{equation}\label{c3a03}
\frac{\nu}{\Omega}\sim\epsilon
\end{equation}
\noindent 	Physically, Eq.~(\ref{c3a03}) implies that, when the plasma is very close to the steady-state [see condition i)], the time derivative in Eq.~(\ref{c3a0}) is considerably smaller than the transit frequency. 
\begin{equation}\label{c3a02}
\frac{\partial}{\partial t}\sim \epsilon^2\frac{V_T}{L_M}
\end{equation}
\noindent Eq.~(\ref{c3a03}) is synonymous with $\nu/\Omega\sim O(\epsilon)$ and admits as a special case $\nu/\Omega\ll \epsilon$. Finally, the fundamental expansion parameter is that of the smallness of the Larmor radius  $\rho_L$ compared with the macroscopic scale length $L_M$. Hence, the problem boils down to seek a solution for $X^{\alpha\mu}$, which may be written as a perturbation expansion in powers of $\epsilon$,
\begin{equation}\label{c3a1}
X^{\alpha\mu}=X^{\alpha\mu}_0+\epsilon X^{\alpha\mu}_1+\epsilon^2X^{\alpha\mu}_2+\cdots
\end{equation}
\noindent where $X^{\alpha\mu}_0$ denotes the {\it reference state}. The definition of this state is given below. From here on, we are tempted to start with the standard perturbation theory. But this approach induces singularities. Indeed, some of the contributions to the correction terms may be proportional to the positive powers of time, $t^n$. It follows that, in spite of the smallness of the parameter, $\epsilon^m$ for long enough times, these terms become very large and invalidate the classification Eq.~(\ref{c3a1}). This fact is so much more troublesome considering that we are interested in a solution close to the steady-state i.e., in the {\it long-time behaviour} of the solution. This difficulty may be overcome by using the multiple time-scale perturbation expansion (see, for example, the book cited in Ref. \cite{davidson}). Thus, we artificially extend the unique time variable $t$ into several dimensionless variables $\varrho_{-1}$, $\varrho_0$, $\varrho_1$, $\dots$, defined by the differential equations 
\begin{equation}\label{c3a2}
\tau_\alpha\ {\dot\varrho_{-1}}=\frac{1}{\varepsilon},\qquad \tau_\alpha\ {\dot\varrho_0}=1,\qquad 
\tau_\alpha\ {\dot\varrho_1}=\epsilon,\ \cdots
\end{equation}
\noindent Once the multiple time-scale solution has been obtained in this form to the desired degree of accuracy, we go back to the original time variable by the replacements
\begin{equation}\label{c3a3}
\varrho_{-1}\rightarrow\frac{1}{\epsilon} \frac{t}{\tau_\alpha},\qquad \varrho_0\rightarrow \frac{t}{\tau_\alpha},\qquad 
\varrho_1\rightarrow \epsilon \frac{t}{\tau_\alpha},\ \cdots
\end{equation}
\noindent The time derivative becomes
\begin{equation}\label{c3a4}
\tau_\alpha\frac{d}{dt}\rightarrow\frac{1}{\epsilon}\frac{d}{d\varrho_{-1}}+\frac{d}{d\varrho_0}+\epsilon\frac{d}{d\varrho_{1}}+\cdots
\end{equation}
\noindent The "singularity" introduced by the term $1/\epsilon$ represents the evolution on the shortest time-scale whereas the dependance on $\varrho_0$ represents the slower evolution on a longer time-scale, etc. Finally, the solution $X^{\alpha\mu}(t)$ is split into many other new variables $X_i^{\alpha\mu}(\varrho_{-1}, \varrho_0,\varrho_1,\cdots)$ (with $i=0,1,2,\cdots$) depending on many (dimensionless) times $\varrho_{-1}, \varrho_0,\varrho_1,\cdots$. As mentioned, the idea is based on the observation that, with the ordering $\rho_L\ll\lambda_{mfp}$ and $\rho_L\ll L_M$, there exist widely separated time-scales in the system. This allows to define the {\it reference state}  $X^{\alpha\mu}_0$ as the state depending only on the {\it shortest time-scale} typically, the inverse Larmor frequency in our problem. Hence, the reference state satisfies the equations $dX_0^{\alpha\mu}/d\varrho_i=0$ (for $i=0,1,\cdots$). The other variables $X^{\alpha\mu}_i$ (with $i=0,1,\cdots$) may, however, depend on a longer time-scales. The unknowns $X_i^{\alpha\mu}$ (with $i=-1,0, 1,\cdots$) are completely determined by imposing the Fredholm solvability (or, the solvability conditions for differential equations) on each iteration. This formal procedure allows to eliminate the secular terms in each order in $\epsilon$ and expansion Eq.~(\ref{c3a1}) is uniformly valid for all times because it takes into account not only the size of the terms $X_i^{\alpha\mu}$, but also their rate of changes in time. According to Eq.~(\ref{c3a00}), on the time-scale $\varrho_2$, the evolution is so slow that we can consider the system at the {\it steady-state}, $X_{st.state}^{\alpha\mu}$, defined by the equations $dX_{st.state}^{\alpha\mu}/d\varrho_i=0$ (for $i=-1, 0,1,\cdots$). Our goal is to examine the evolution of the system towards the steady-state by stopping calculations at the time derivative $\epsilon d/d\varrho_{1}$. Note that this mathematical ruse is widely used in literature, especially for solving the kinetic equations, expressed in natural guiding centre variables, valid within the drift approximation (see, for example \cite{balescu2}).

\noindent A quite similar procedure can be found in Ref. \cite{rosenbluth} for solving the drift-kinetic equation, $f$, for the toroidal rotation induced by neutral beam injection. In this case, the distribution function for fast ions is expanded in powers of the small parameter entering in the problem (the ratio of the slowing down rate to the bounce frequency) and the "singularity"  $1/\epsilon$ is contained in the term at the lowest order, $f_{-1}$, and not in the time derivative. Also in this approach, the differential equation for the "reference state" (i.e. for $f_{-1}$) is determined to the dominant order. The evolution on the various time-scales is obtained by applying the bounce averaging to the dynamical equations for each order in the perturbation expansion. With this method, the authors have been able "to capture" the main features of the slowly varying functions. The macroscopic observables, like the torque on the plasma, have then been obtained by applying the magnetic surface average.

\noindent Coming back to our problem, condition {\bf i)} is fulfilled if, during the relaxation, Eq.~(\ref{c3a}) is expanded starting from the term
\begin{equation}\label{c3a4}
\tau_\alpha\ {\dot\varrho_{-1}}=\frac{1}{\epsilon}
\end{equation}
\noindent Taking into account Eq.~(\ref{c3a1}) and Eq.~(\ref{c3a3}),  at the lowest orders we find 
\begin{equation}\label{c3c}
\!\!
\left\{ \begin{array}{ll}
\frac{dX_0^{\alpha\mu}}{d\varrho_{-1}}=-\epsilon
g_{0}^{\mu\lambda}N_{0\lambda\kappa}
\Bigl({\tilde g}_0^{\kappa\nu}J_{0\nu}^\alpha-X_0^{\alpha\kappa}\Bigr)&\\
&\\
\frac{dX_1^{\alpha\mu}}{d\varrho_{-1}}+\frac{dX_0^{\alpha\mu}}{d\varrho_{0}}=-
g_{0}^{\mu\lambda}N_{0\lambda\kappa}\Bigl({\tilde g}_0^{\kappa\nu}J_{0\nu}^\alpha-X_0^{\alpha\kappa}\Bigr)&
 \end{array}
\right.
\end{equation}
\noindent with $N_{0\lambda\kappa}\equiv{\tilde M}^\eta_{0\lambda}\delta_{\eta\kappa}$ and ${\tilde M}^\eta_{0\nu}$ is the inverse matrix of $M^\mu_{0\eta}\equiv\delta_\eta^\mu+g_0^{\mu\lambda}X_0^{\alpha\kappa}g_{0\eta\kappa,\lambda}$ i.e., ${\tilde M}^\eta_{0\nu} M^\mu_{0\eta}=\delta^\mu_\nu$.
\noindent We note that Eq.~(\ref{c3c}) transforms in a covariant way under the following thermodynamic coordinate transformations (TCT) \cite{sonninog}
\begin{eqnarray}\label{eq1}
&&X^{'\alpha\mu}=\frac{\partial X^{'\alpha\mu}}{\partial X^{\alpha\nu}} X^{\alpha\nu}\nonumber\\
&& J^{'\alpha}_{\mu}=\frac{\partial X^{\alpha\nu}}{\partial X^{'\alpha\mu}}J_{\nu}^\alpha
\end{eqnarray}
\noindent Indeed, by inserting transformations (\ref{eq1}) into the Eqs~(\ref{c3c}) we find
\begin{equation}\label{eq2}
\!\left\{ \begin{array}{ll}
\frac{dX_0^{'\alpha\mu}}{d\varrho_{-1}}=-\epsilon
{g}_{0}^{\prime\mu\lambda}
\Bigl({\tilde g}_0^{\prime\kappa\nu}J_{0\nu}^{'\alpha}-X_0^{'\alpha\kappa}\Bigr)N_{0\eta\beta}\frac{\partial X^{\alpha\eta}}{\partial X^{'\alpha\lambda}}\frac{\partial X^{\alpha\beta}}{\partial X^{'\alpha\kappa}}&\\
&\\
\frac{dX_1^{'\alpha\mu}}{d\varrho_{-1}}+\frac{dX_0^{'\alpha\mu}}{d\varrho_{0}}=
-{g}_{0}^{\prime\mu\lambda}\Bigl({\tilde g}_0^{\prime\kappa\nu}J_{0\nu}^{'\alpha}-X_0^{'\alpha\kappa}\Bigr)N_{0\eta\beta}\frac{\partial X^{\alpha\eta}}{\partial X^{'\alpha\lambda}}\frac{\partial X^{\alpha\beta}}{\partial X^{'\alpha\kappa}}
&
 \end{array}
\right.
\end{equation}
\noindent or
\begin{eqnarray}\label{eq2a}
&&\left\{ \begin{array}{ll}
\frac{dX_0^{'\alpha\mu}}{d\varrho_{-1}}=-\epsilon
{g}_{0}^{\prime\mu\lambda}N'_{0\lambda\kappa}
\Bigl({\tilde g}_0^{\prime\kappa\nu}J_{0\nu}^{'\alpha}-X_0^{'\alpha\kappa}\Bigr)&\\
&\\
\frac{dX_1^{'\alpha\mu}}{d\varrho_{-1}}+\frac{dX_0^{'\alpha\mu}}{d\varrho_{0}}=-
{g}_{0}^{\prime\mu\lambda}N'_{0\lambda\kappa}\Bigl({\tilde g}_0^{\prime\kappa\nu}J_{0\nu}^{'\alpha}-X_0^{'\alpha\kappa}\Bigr)
 \end{array}
 \right.\\
 \end{eqnarray}
\noindent where
\begin{equation}
N'_{0\lambda\kappa}\equiv N_{0\eta\beta}\frac{\partial X^{\alpha\eta}}{\partial X^{'\alpha\lambda}}\frac{\partial X^{\alpha\beta}}{\partial X^{'\alpha\kappa}}
\end{equation}
\noindent Hence, matrix $N_{0\lambda\kappa}$ transforms like a thermodynamic tensor of second rank. The TCT are the most general forces-transformations leaving invariant the expression of the entropy production and the Glansdorff-Prigogine dissipative quantity ${\mathcal P}$ defined as 
\begin{equation}\label{eq3}
{\mathcal P}\equiv J^\alpha_\mu\frac{dX^{\alpha\mu}}{dt}
\end{equation}
\noindent Notice that from Eq.(\ref{eq1}), the following important identities are easily derived \cite{sonninog}
\begin{equation}\label{tct2}
X^{\alpha\nu}\frac{\partial^2X'^{\alpha\mu}}{\partial X^{\alpha\nu}\partial X^{\alpha\kappa}}=0\qquad ;\qquad X'^{\alpha\nu}\frac{\partial^2X^{\alpha\mu}}{\partial X'^{\alpha\nu}\partial X'^{\alpha\kappa}}=0
\end{equation}
\noindent So the linear transformations 
\begin{equation}\label{tct1a}
X'^{\alpha\mu}=c_\nu^{\alpha\mu} X^{\alpha\nu}
\end{equation}
\noindent where $c_\nu^{\alpha\mu}$ are constant coefficients (i.e., independent of the thermodynamic forces) are a special, but very important, class of the TCT. The covariance under translation of the variables $X^{\alpha\nu}$ has been widely used in literature. We cite, as an example, the work of Hinton and Hazeltine \cite{hinton} on the interpolation-expression (i.e., the expression valid for the fully collisional as well as the low-collisional transport regime) for the diffusion coefficient of magnetically confined collisional-plasmas. The original interpolation formula, found by the authors by solving the gyrokinetic equations, showed a wrong sign as one goes from the low-collisional regime to the fully collisional regime. Hinton and Hazeltine have been able to restore the correct sign by performing a linear transformation of the thermodynamic forces. The linear transformations constitute a very important class of the TCT and it is always possible to associate to them a clear physical interpretation. For example, in the case of the Hinton-Hazeltine interpolation formula, the transformed flux corresponds to the {\it total averaged radial energy flux} obtained by the linear combination of the two original fluxes, the {\it heat flux} and the {\it enthalpy carried out by the particle flux}. However as already mentioned, the balance equations, as well as the closure equations, are covariant under the entire class of the TCT and, {\it a priori}, there are no reasons for retaining only the linear transformations.

\noindent ${\mathcal P}$ can be obtained by deriving both sides of the first equation of Eqs~(\ref{c3c}) with respect to parameter $\varrho_{-1}$. However, in accordance with the request {\bf ii)} mentioned in section \ref{plasma}, this operation should be performed in such a way to preserve the covariance under TCT. For this, we undertake the covariant differentiation along a curve of both sides of the first equation of Eqs~(\ref{c3c}) \cite{sonninog} 
\begin{align}\label{eq4}
\!\!\!\!\!\!\!
\frac{d^2X_0^{\alpha\mu}}{d\varrho_{-1}^2}+\Gamma_{\lambda\kappa}^\mu\frac{dX_0^{\alpha\lambda}}{d\varrho_{-1}}\frac{dX_0^{\alpha\kappa}}{d\varrho_{-1}}
=&\ \epsilon^2 \Bigl({\tilde g}_0^{\iota\nu}J_{0\nu}^\alpha-X_0^{\alpha\iota}\Bigr)
\Bigl({\tilde g}_0^{{\kappa}'{\nu}'}J_{0{\nu}'}^\alpha-X_0^{\alpha{\kappa}'}\Bigr) g_0^{\lambda\eta}N_{0\eta{\kappa'}}\\
&\ {\rm x}\quad \Bigl[\Bigl({g}_0^{\mu{\eta}'}N_{0{\eta}'\iota}\Bigr)_{,\lambda}
+\Gamma^\mu_{0\lambda\kappa}g_0^{\kappa{\eta}'}N_{0{\eta}'\iota}
\Bigr]\nonumber
\end{align}
\noindent where comma $(,)$ stands for partial differentiation with respect to the thermodynamic forces and $\Gamma_{0\lambda\kappa}^\mu$ is the affine connection evaluated at the lowest order. The elements $\Gamma_{\lambda\kappa}^\mu$ transform like an affine connection under the coordinate transformations Eqs~(\ref{eq1}). Its expression is reported in Ref. \cite{sonninog} 
\begin{equation}\label{eq5}
\Gamma_{\lambda\kappa}^\mu=\frac{1}{2}g^{\mu\nu}\Bigl(\frac{\partial g_{\lambda\nu}}{\partial X^{\alpha\kappa}}+\frac{\partial g_{\nu\kappa}}{\partial X^{\alpha\lambda}}-\frac{\partial g_{\lambda\kappa}}{\partial X^{\alpha\nu}}\Bigr)+\frac{1}{2\sigma^\alpha}X^{\alpha\mu} X^{\alpha\nu}\frac{\partial g_{\lambda\kappa}}{\partial X^{\alpha\nu}}
\end{equation}
\noindent From Eq.~(\ref{c3a1}), we obtain
\begin{equation}\label{c5}
\mathcal{U}^{\alpha\mu} (X,\varrho_{-1})\equiv \frac{d^2X^{\alpha\mu}}{d\varrho_{-1}^2}+\Gamma^\mu_{\lambda\kappa}\frac{dX^{\alpha\lambda}}{d\varrho_{-1}}\frac{dX^{\alpha\kappa}}{d\varrho_{-1}}\sim O(\epsilon^2)
\end{equation}
\noindent This can be easily seen by noting that
\begin{equation}\label{c5n}
\mathcal{U}^{\alpha\mu} (X,\varrho_{-1})=\mathcal{U}^{\alpha\mu} (X_0,\varrho_{-1})
+\epsilon\ \!{\mathcal H}^{\alpha\mu}(X_0,X_1,\varrho_{-1})+O(\epsilon^2)
\end{equation}
\noindent where
\begin{equation}\label{c5a}
{\mathcal H}^{\alpha\mu}(X_0,X_1,\varrho_{-1})\equiv\frac{d^2X_1^{\alpha\mu}}{d\varrho_{-1}^2}+\Gamma^\mu_{0\lambda\kappa}\Bigl(
\frac{dX_0^{\alpha\lambda}}{d\varrho_{-1}}\frac{dX_1^{\alpha\kappa}}{d\varrho_{-1}}+\frac{dX_1^{\alpha\lambda}}{d\varrho_{-1}}\frac{dX_0^{\alpha\kappa}}{d\varrho_{-1}}\Bigr)
\end{equation}
\noindent From Eqs~(\ref{c3c}) we get
\begin{equation}\label{eq6}
{\mathcal H}^{\alpha\mu}(X_0,X_1,\varrho_{-1})\sim O(\epsilon)
\end{equation}
\noindent Thus, taking into account Eq.~(\ref{eq4}), close to the steady-state, the evolutionary-trajectory (ev.-tr.) of the system satisfies, at the dominant order, the equation
\begin{equation}\label{eq7}
\mathcal{U}^{\alpha\mu} (X_{ev.-tr.},\varrho_{-1})=0+O(\epsilon^2)
\end{equation}
\noindent We can check that Eq.~(\ref{eq7}) satisfies condition {\bf ii)}. Indeed, multiplying this equation with the thermodynamic forces $X^{\alpha\mu}$ and contracting, we obtain
\begin{equation}\label{c5a}
J^\alpha_\mu\frac{d^2X_{ev.-tr.}^{\alpha\mu}}{d\varrho_{-1}^2}+X_{ev.-tr.}^{\alpha\mu}g_{\mu\nu,\kappa}\frac{dX_{ev.-tr.}^{\alpha\nu}}{d\varrho_{-1}} \frac{dX_{ev.-tr.}^{\alpha\kappa}}{d\varrho_{-1}} =0+ O(\epsilon^2)
\end{equation}
\noindent Recalling that $g_{\mu\nu}\frac{dX_{ev.-tr.}^\mu}{d\varsigma} \frac{dX_{ev.-tr.}^\nu}{d\varsigma} =1$, where $\varsigma$ indicates the {\it arc-parameter} \cite{sonninog}, we get the following identity
\begin{equation}\label{c5b}
 J^\alpha_\mu\frac{d^2X_{ev.-tr.}^{\alpha\mu}}{d\varrho_{-1}^2}=\frac{d{\tilde P}}{d\varrho_{-1}}-\Bigl(\frac{d\varsigma}{d\varrho_{-1}}\Bigr)^2-X_{ev.-tr.}^{\alpha\mu}g_{\mu\nu,\kappa}
 \frac{dX_{ev.-tr.}^{\alpha\nu}}{d\varrho_{-1}} \frac{dX_{ev.-tr.}^{\alpha\kappa}}{d\varrho_{-1}} 
\end{equation}
\noindent where ${\tilde P}=J^\alpha_\mu \frac{dX_{ev.-tr.}^{\alpha\mu}}{d\varrho_{-1}} $. Inserting Eq.~(\ref{c5b}) into Eq.~(\ref{c5a}), we obtain, at the leading order, the expression
\begin{equation}\label{c5c}
\frac{d {\tilde P}}{d\varrho_{-1}}=\Bigl(\frac{d\varsigma}{d\varrho_{-1}}\Bigr)^2+O(\epsilon^2)
\end{equation}
\noindent Integrating from the initial condition to the steady-state, we get
\begin{equation}\label{c5d}
{\tilde P}(X^{\alpha}_{st.state})-{\tilde P}=\int\Bigl(\frac{d\varsigma}{d\varrho'_{-1}}\Bigr)^2d\varrho'_{-1}\ge0
\end{equation}
\noindent At the steady-state we have ${\dot\varrho_{-1}}{\tilde P}(X^{\alpha}_{st.state})=0$, so we finally obtain
\begin{equation}\label{c5e}
{\mathcal P}={\dot\varrho_{-1}}{\tilde P}={\dot\varrho_{-1}}J^\alpha_\mu\frac{dX_{ev.-tr.}^{\alpha\mu}}{d\varrho_{-1}}=-{\dot\varrho_{-1}}\int\Bigl(\frac{d\varsigma}{d\varrho'_{-1}}\Bigr)^2d\varrho'_{-1}\le0
\end{equation}
\noindent where the inequality is only saturated at the steady-state. Thus, after fast transient, up to $\epsilon^2$, the correct (perturbed) evolution equations close to the steady-state are
\begin{eqnarray}\label{c6a}
\left\{ \begin{array}{ll}
\frac{d^2X_0^{\alpha\mu}}{d\varrho_{-1}^2}+\Gamma_{\lambda\kappa}^\mu\frac{dX_0^{\alpha\lambda}}{d\varrho_{-1}}\frac{dX_0^{\alpha\kappa}}{d\varrho_{-1}}=0&\\
&\\
\frac{dX_1^{\alpha\mu}}{d\varrho_{-1}}=-\frac{dX_0^{\alpha\mu}}{d\varrho_{0}}-
g_{0}^{\mu\lambda}N_{0\lambda\kappa}\Bigl({\tilde g}_0^{\kappa\nu}J_{0\nu}^\alpha-X_0^{\alpha\kappa}\Bigr)&
 \end{array}
\right.
\end{eqnarray}
\noindent Eqs.~(\ref{c6a}) are covariant under the TCT, satisfy the UCE and are derived from the balance equations Eqs~(\ref{c3a}). This set of equations are {\it closed} by recalling the definition of the reference state. Hence, $dX_0^{\alpha\mu}/d\varrho_{0}=0$. So, near the steady-states, the equations describing the relaxation of magnetically confined plasmas read (see also the footnote \footnote{Note that, from the mathematical point of view, the condition $dX_0^{\alpha\mu}/d\varrho_{0}=0$ ensures the solvability conditions for Eqs~(\ref{c6a}). \label{fredholm}})
\begin{eqnarray}\label{c6b}
\left\{ \begin{array}{ll}
\frac{d^2X_0^{\alpha\mu}}{d\varrho^2}+\Gamma_{\lambda\kappa}^\mu\frac{dX_0^{\alpha\lambda}}{d\varrho}\frac{dX_0^{\alpha\kappa}}{d\varrho}=0&\\
&\\
\frac{dX_1^{\alpha\mu}}{d\varrho}=-
g_{0}^{\mu\lambda}N_{0\lambda\kappa}\Bigl({\tilde g}_0^{\kappa\nu}J_{0\nu}^\alpha-X_0^{\alpha\kappa}\Bigr)&
 \end{array}
\right.
\end{eqnarray}
\noindent where, for simplicity, we have suppressed in the parameter $\varrho_{-1}$ the sub-index "$-1$". System (\ref{c6b}) admits a unique solution satisfying the initial conditions $ X_0^{\alpha\mu}(0)=X^{\alpha\mu}(0)$ [i.e., $X_1^{\alpha\mu}(0)=0$] and $X_{0\ st.state}^{\alpha\mu}+\epsilon X_{1\ st.state}^{\alpha\mu}=X_{st.state}^{\alpha\mu}$. 

\noindent It is known that, by a suitable choice of the parameter $\varrho$, the differential equation for the shortest path in the thermodynamic space reads (see, for example, Ref. \cite{eisenhart})
\begin{equation}\label{c6}
{\mathcal U}^{\alpha\mu} (X_{s.path},\varrho)\equiv
\frac{d^2X_{s.path}^{\alpha\mu}}{d\varrho^2}+\Gamma^\mu_{\lambda\kappa}\frac{dX_{s.path}^{\alpha\lambda}}{d\varrho}\frac{dX_{s.path}^{\alpha\kappa}}{d\varrho}=0
\end{equation}
\noindent Hence, from Eqs~(\ref{eq7}) we straightforwardly obtain the following result: {\it In the collisional transport regimes and under the validity of the drift parameter approximation, after fast transients, tokamak-plasmas relax to stable steady-states in the generalized frictions space so that}
\begin{equation}\label{c8a}
\begin{split}
&\!\!\!\!\!\!\!\!\!\!\!\!\!\!\!\!\!\!\!\!\!\!\!\!\!\!\!\!\!\!\!\!
J^\alpha_\mu\mathcal{U}^{\alpha\mu} (X_{ev.-tr.},\varrho) - J^\alpha_\mu{\mathcal U}^{\alpha\mu} (X_{s.path},\varrho)=O(\epsilon^2)\\
&\!\!\!\!\!\!\!\!\!\!\!\!\!\!\!\!\!\!\!\!\!\!\!\!\!\!\!\!\!\!\!\!
{\rm (no\ summation\ over\ }\alpha{\rm)}
\end{split}
\end{equation}
\noindent It is worthwhile mentioning that Eq.~(\ref{c8a}) can be derived from Eq.~(\ref{eq7}), but the reverse is not true. Indeed, in our example we made use of the drift parameter approximation. In this particular case, parameter $\epsilon$ is independent of the distance of the system to the steady-state and we obtained the result that, at the first order of approximation, collisional plasmas relax to the stationary states along the shortest path traced out in the  space of the parallel generalized frictions. However, this result is not satisfied in general. The small parameter $\epsilon$, entering in a problem of relaxation depends , in general, on the distance of the system to the steady-state. It is possible to show that Eq.~(\ref{c8a}) remains valid also when $\epsilon$ depends on the distance to the steady-state. In the paper reported in Ref.~ \cite{sonninojmp}, several examples of relaxation of chemical reactions to stable steady-states in the chemical affinities space are analyzed.  Even in these cases, it is possible to show the validity  of Eq.~(\ref{c8a}) when the chemical systems relax near the stationary states. In these examples, $X^\mu$ and $J_\mu$ are the chemical affinities and the chemical reaction rates, respectively and $\epsilon$ measures the distance of the system from the stationary state. 

\noindent Note that in Onsager's region, $\Gamma_{\lambda\kappa}^\mu\rightarrow0$ and $(g_0^{\mu\kappa}N_{0\kappa\nu})_{,\lambda}\rightarrow0$. Moreover, $\varrho\propto\varsigma$ (see Ref. \cite{sonninog}). Hence, the evolution equation for $X_0^{\alpha\mu}$ reduces to [see Eq.~(\ref{eq4})]
\begin{equation}\label{c9}
\frac{d^2X_0^{\alpha\mu}}{d\varsigma^2}=O(\epsilon^2)\quad {\rm or} \quad 
X_0^{\alpha\mu}(\varsigma)=X_{0\ st.state}^{\alpha\mu}+\Bigl(1-\frac{\varsigma}{l}\Bigr)\Bigl[X^{\alpha\mu}(0)-X_{0\ st.state}^{\alpha\mu}\Bigr]+O(\epsilon^2)
\end{equation}
\noindent In other words, in the Onsager region, at the dominant order, the solution is a straight line in the space spanned by the parallel generalized frictions. 

\subsection{Spatially Dependent Magnetically Confined Plasmas}\label{plasma2}

\noindent In this case, firstly  we have to develop the space-time dependent thermodynamic forces (${\mathcal X}^{\alpha\mu}$), flows (${\mathcal J}_\mu^\alpha$) and transport coefficients in (spatial) Fourier's series. Then, we perform the same calculations as in the homogeneous case by taking into account the {\it slaving principle} \cite{haken} (see the footnote \footnote{The slaving principle establishes that in a relaxation process, contributions from different wave-numbers are negligible with respect to those with same wave-numbers.\label{slaving}}). Close to the steady-state, we have \cite{sonninog}
\begin{eqnarray}\label{c10}
&&\int_V {\mathcal J}^\alpha_\mu\frac{d{\mathcal X}^{\alpha\mu}}{d\varrho}dv\simeq V{\widehat J}^\alpha_{{(\bf 0)}\mu}\frac{d{\widehat X}^{\alpha\mu}_{({\bf 0})}}{d\varrho}\leq 0
\qquad\qquad\quad {\rm with}\nonumber\\
&&
{\widehat J}^\alpha_{\mu ({\bf k})}(t)=\frac{1}{V}\int_V {\mathcal J}^\alpha_\mu({\bf r},t)
\exp(-i{\bf k}\cdot{\bf r})dv\nonumber\\
&&
{\widehat X}^{\alpha\mu}_{({\bf k}')}(t)=\frac{1}{V}\int_V{\mathcal X}^{\alpha\mu}({\bf r},t)
\exp(-i{\bf k}'\cdot{\bf r})dv
\end{eqnarray}
\noindent $dv$ denotes a (spatial) volume element of the system, and the integration is over the entire space $V$ occupied by the system in question. The evolution equation for ${\widehat J}^\alpha_{{(\bf k)}\mu}$ at the {\it thermodynamic mode} (i.e., the mode with wave-number ${\bf k} = {\bf 0}$), reads
\begin{equation}\label{c11}
\tau_\alpha{\dot {\widehat J^\alpha}_{({\bf 0})\mu}}=-
\delta_{\mu\kappa}\bigl({\widehat {\tilde g}}_{({\bf 0})}^{\kappa\nu}{\widehat J}^\alpha_{{({\bf 0})}\nu}-{\widehat X}_{({\bf 0})}^{\alpha\kappa}\bigr)+{\rm O.T.}({\bf k},{\bf k'})_{{\bf k}\neq{\bf k'}\neq {\bf 0}}
+O(\epsilon^2)
\end{equation}
\noindent where ${\rm O.T.}({\bf k},{\bf k'})_{{\bf k}\neq{\bf k'}\neq {\bf 0}}$ stands for {\it other contributions from different wave-numbers} and ${\widehat g}_{({\bf 0})}^{\kappa\nu}$ is a (time-dependent) positive definite matrix \cite{sonninog}. Due to the slaving principle, near the steady-state, the evolution equations for ${\widehat J}^\alpha_{{(\bf 0)}\mu}$ reduce to
\begin{equation}\label{c12}
\tau_\alpha{\dot {\widehat J^\alpha}_{({\bf 0})\mu}}\simeq-
\delta_{\mu\kappa}\bigl({\widehat {\tilde g}}_{({\bf 0})}^{\kappa\nu}{\widehat J}^\alpha_{{({\bf 0})}\nu}-{\widehat X}_{({\bf 0})}^{\alpha\kappa}\bigr)+O(\epsilon^2)
\end{equation}
\noindent By performing now the same multiple time-scale calculations as for the homogeneous case, we arrive to the following final result: {\it the thermodynamic mode} (${\bf k} = {\bf 0}$) {\it relaxes to the steady-state according to the law}
\begin{equation}\label{c13}
{\widehat J}^\alpha_{({\bf 0})\mu}\mathcal{U}^{\alpha\mu} ({\widehat X}_{({\bf 0})},\varrho) - {\widehat J}^\alpha_{({\bf 0})\mu}{\mathcal U}_{s.path}^{\alpha\mu} ({\widehat X}_{({\bf 0})},\varrho)=O(\epsilon^2)
\end{equation}
\vskip 0.2truecm
\section{Conclusions}\label{cs}
\vskip 0.2truecm

In this concluding section, we shall use the results previously obtained for finally deriving the expression of the affine connection. The Universal Criterion of Evolution theorem has been demonstrated by Glansdorff and Prigogine through the balance equations. From this theorem and by analyzing examples of relaxation of chemical systems close to steady-states, we may obtain the expression for the affine connection of the thermodynamical forces space. This construction may be made "step by step".  We multiply the first equation of Eqs~(\ref{c3c}) by $g_{\mu\nu}$ and sum over the index $\mu$. By taking the derivative, with respect to parameter $\varrho$ ($\varrho=\varrho_{-1}$), of both sides of the resulting equation, we find (for simplicity, we omit the index $\alpha$)
\begin{equation}\label{concl1}
\begin{split}
&g_{\mu\nu}\frac{d^2X^\mu}{d\varrho^2}+\frac{1}{2}\bigl(g_{\nu\lambda,\kappa}+g_{\nu\kappa,\lambda}\bigr)\frac{dX^\lambda}{d\varrho}\frac{dX^\kappa}{d\varrho}\\
&=\epsilon^2 \Bigl({\tilde g}_0^{\iota\nu}J_{0\nu}^\alpha-X_0^{\alpha\iota}\Bigr)
\Bigl({\tilde g}_0^{{\kappa}'{\nu}'}J_{0{\nu}'}^\alpha-X_0^{\alpha{\kappa}'}\Bigr)g_0^{\lambda\eta}N_{0\eta{\kappa'}}N_{0{\nu}\iota,\lambda}
\end{split}
\end{equation}
\noindent Thus, close to the steady-state, we have
\begin{equation}\label{concl3}
g_{\mu\nu}\frac{d^2X^\mu}{d\varrho^2}+\frac{1}{2}\bigl(g_{\nu\lambda,\kappa}+g_{\nu\kappa,\lambda}\bigr)\frac{dX^\lambda}{d\varrho}\frac{dX^\kappa}{d\varrho}=O(\epsilon^2)
\end{equation}
\noindent At the lowest order, previous equation may be rewritten as
\begin{equation}\label{concl3a}
\frac{d^2X^\mu}{d\varrho^2}+\frac{1}{2}g^{\mu\nu}\bigl(g_{\nu\lambda,\kappa}+g_{\nu\kappa,\lambda}\bigr)\frac{dX^\lambda}{d\varrho}\frac{dX^\kappa}{d\varrho}=O(\epsilon^2)
\end{equation}
\noindent It is not difficult to show that Eq.~(\ref{concl3a}) satisfies the UCE but it is {\it not} covariant under the TCT. This because the term $1/2g^{\mu\nu}\bigl(g_{\nu\lambda,\kappa}+g_{\nu\kappa,\lambda}\bigr)$, does not transform as an affine connection under the TCT. This condition is satisfied by adding the Levi-Civita term $-1/2g^{\mu\nu}g_{\lambda\kappa,\nu}$. Now, if one wants the Universal Criterion of Evolution satisfied also when the system relaxes along a shortest path, without imposing a priori {\it any} restrictions on transport coefficients, an extra term to the Levi-Civita affine connection should be added. It can be checked that the most general expression for this extra term is $1/(2\sigma)X^\mu X^\nu g_{\lambda\kappa,\nu}$ \cite{sonninog}. Hence, the affine connection my be written as
\begin{equation}\label{e1}
{\tilde\Gamma}^\mu_{\lambda\kappa}=\frac{1}{2}g^{\mu\nu}\Bigl(\frac{\partial g_{\nu\kappa}}{\partial X^\lambda}+\frac{\partial g_{\lambda\nu}}{\partial X^\kappa}-\frac{\partial g_{\lambda\kappa}}{\partial X^\nu}\Bigr)+\frac{1}{2\sigma}X^\mu X^\nu\frac{\partial g_{\lambda\kappa}}{\partial X^\nu}
\end{equation}
\noindent However, from the general theory on non-Riemannian geometry, we know that both equations
\begin{equation}\label{e1a}
\frac{d^2X^\mu}{d\varrho^2}+{{\tilde\Gamma}}^\mu_{\lambda\kappa}\frac{dX^\lambda}{d\varrho}\frac{dX^\kappa}{d\varrho}=0
\end{equation}
\noindent and
\begin{eqnarray}\label{e1b}
&&\frac{d^2X^\mu}{dp^2}+\Gamma^\mu_{\lambda\kappa}\frac{dX^\lambda}{d p}\frac{dX^\kappa}{dp}=0\qquad {\rm with}\nonumber\\
&& \Gamma^\mu_{\lambda\kappa}={{\tilde\Gamma}}^\mu_{\lambda\kappa} +\delta^\mu_{\lambda}\psi_\kappa+
\delta^\mu_{\kappa}\psi_\lambda
\end{eqnarray}
\noindent are evolution equations for the {\it same} shortest path \cite{eisenhart}. In Eq.~(\ref{e1b}), $p$ is a parameter, related to $\varrho$ by
\begin{equation}\label{e1a1}
\frac{d\varrho}{dp}=c\exp\Bigl(-2\int\psi_\nu dX^\nu\Bigr)\ ;\quad c={\rm positive\ const.}
\end{equation}
\noindent and $\psi_\nu$ is an arbitrary vector under TCT. In literature, modifications of the connection similar to Eqs~(\ref{e1b}) is referred to as {\it projective transformations} of the connection and $\psi_\kappa$ as the {\it projective covariant vector}. Therefore, the UCE is satisfied for every shortest path constructed with affine connections $\Gamma_{\lambda\kappa}^\mu$, linked to ${\tilde\Gamma}_{\lambda\kappa}^\mu$ by projective transformations Eqs~(\ref{e1b}) \cite{eisenhart}. By standard methods in non-Riemannian geometry, it is not difficult to determine the final expression of the projective-invariant affine connection \cite{sonninog}, \cite{eisenhart}:
\begin{equation}\label{e2}
\begin{split}
\!\!\!\!
\Gamma^\mu_{\lambda\kappa}=&
\frac{1}{2}g^{\mu\nu}\Bigl(\frac{\partial g_{\lambda\nu}}{\partial X^\kappa}+\frac{\partial g_{\kappa\nu}}{\partial X^\lambda}-\frac{\partial g_{\lambda\kappa}}{\partial X^\nu}\Bigr)
+\frac{X^\mu X^\nu}{2\sigma}\frac{\partial g_{\lambda\kappa}}{\partial X^\nu}-
\frac{X^\eta X^\nu}{2(n+1)\sigma}\Bigl[
\delta^\mu_\lambda \frac{\partial g_{\kappa\eta}}{\partial X^\nu}
+\delta^\mu_\kappa \frac{\partial g_{\lambda\eta}}{\partial X^\nu}\Bigr]
\end{split}
\end{equation}
\noindent where $n$ denotes the dimension of the generalized frictions space. We can check that the expression reported in Eq.~(\ref{e2}) transforms under TCT (therefore, also under the translation of the variables $X^{\alpha\mu}$) as an affine connection. A non-Riemannian geometry has been successively constructed out of the components of the affine connections \cite{sonninog}. The main conclusion of the present analysis is thus:

\noindent  {\it Close to steady-states, the geometry of the thermodynamic space is non-Riemannian with affine connection given by Eq.~(\ref{e2})}. The knowledge of the expression for the affine connection is a fundamental prerequisite for the construction of the (nonlinear) closure theory on transport processes. In Ref.~\cite{sonninog}, the curvature tensor and the nonlinear transport equations have successively been derived from Eq.~(\ref{e2}) and by introducing the following assumption:

\noindent {\it There exists a thermodynamic action, scalar under $TCT$, which is stationary with respect to arbitrary variations in the transport coefficients and the affine connection}. 

\noindent From this principle, a set of closure equations, constraints, and boundary conditions have been derived. These equations determine the nonlinear corrections to the linear ("Onsager") transport coefficients. The nonlinear transport equations have successively been used for computing the particle and energy losses in magnetically confined plasmas and for deriving the distribution density functions for the species $\alpha$ in several collisional transport regimes \cite{sonnino3} and \cite{sonnino4}. We already mentioned in the introduction that, the thermodynamic forces are the variables, which characterize the departure of the system from the (thermodynamic) equilibrium. As a consequence, for plasmas not sufficiently close to the equilibrium, the distribution function for the fluctuations of the thermodynamic quantities deviates from a Maxwellian and, in general, the closure equations may not be linearized with respect to the thermodynamic forces. From the microscopic point of view, this aspect can be easily understood in the following way. Let us indicate with $\{Y_\mu\} $ a set of $n$ extensive variables describing the state of the plasma. We shall write the distribution function $f(Y)$ for which $\{Y_\mu\} $ lies between $\{Y_\mu\} $ and $\{Y_\mu\}+\{dY_\mu\}$ as (see also the footnote \footnote{From a more fundamental point of view, the validity of Eq.~(\ref{cd1}) implies that on a kinetic model the extensive variables $\{Y_\mu\}$ are algebraic sums of microscopic variables \cite{degroot}. \label{fluctuation}})
\begin{eqnarray}\label{cd1}
f(Y)dY
&&\equiv f(Y_1,Y_2,\cdots , Y_n)dY_1dY_2\cdots dY_n
\nonumber\\
&&={\hat c} \exp\Bigl\{-\frac{h^{\mu\nu}(Y)[Y_\nu-{\bar Y}_\nu][Y_\mu-{\bar Y}_\mu]}{2K_B}\Bigr\}dY_1dY_2\cdots dY_n
\end{eqnarray}
\noindent where ${\hat c}$ is the {\it normalization constant}, $K_B$ Boltzmann's constant and $h^{\mu\nu}$ the elements of a symmetric positive matrix, respectively. ${\bar Y}_\mu$ denote the mean values of $Y_\mu$
\begin{equation}\label{cd2}
{\bar Y_\mu}=\int f(Y_1,Y_2,\cdots , Y_n) Y_\mu dY_1dY_2\cdots dY_n
\end{equation}
\noindent Hence, the fluctuations $\xi_i$ defined as 
\begin{equation}\label{cd3}
\xi_\mu\equiv Y_\mu-{\bar Y}_\mu\qquad{\rm with}\quad (\mu=1,\cdots , n)
\end{equation}
\noindent obey the distribution function
\begin{equation}\label{cd4}
f(\xi_1,\xi_2,\cdots \xi_n)= {\hat c}\exp\Bigl[-\frac{h^{\mu\nu}(\xi)\xi_\nu\xi_\mu}{2K_B}\Bigr]
\end{equation}
\noindent The following quantities will now be introduced 
\begin{equation}\label{cd5}
X^\mu\equiv K_B\frac{\partial \log f}{\partial\xi_\mu}
\end{equation}
\noindent From this definition we find
\begin{equation}\label{cd6}
X^\mu=-h^{\mu\nu}\xi_\nu-\frac{1}{2}
\frac{\partial h^{\kappa\nu}}{\partial\xi_\mu}\xi_\kappa\xi_\nu
\end{equation}
\noindent The variables $X^\mu$, referred to as the {\it thermodynamic forces}, clearly characterize the departure of the system from the equilibrium (notice that $X^\mu\rightarrow0$ as $\xi_\mu\rightarrow0$, for $\mu=1,\cdots n$). Near equilibrium, the matrix $h^{\mu\nu}$ is close to a constant (symmetric positive) matrix, say $h_0^{\mu\nu}$, and the distribution function, Eq.~(\ref{cd1}), takes the form of a Maxwellian. Thus, we can write
\begin{eqnarray}\label{cd7}
&&X^\mu\simeq-h_0^{\mu\nu}\xi_\nu\nonumber\\
&& f(\xi)\simeq{\hat c}\exp\Bigl[-\frac{h_0^{\mu\nu}\xi_\nu\xi_\mu}{2K_B}\Bigr]
\end{eqnarray}
\noindent for thermodynamic systems close to equilibrium. The {\it conjugate thermodynamic flow} are defined as 
\begin{equation}\label{cd8}
J_\mu\equiv\frac{d\xi_\mu}{dt}
\end{equation}
\noindent We shall now assume that decay of a fluctuation follows the ordinary macroscopic law (see, for example, Ref.~\cite{sonninojmp}) and write
\begin{equation}\label{cd9}
\frac{d\xi_\mu}{dt}=-k_\mu^\nu (\xi)\xi_\nu
\end{equation}
\noindent Near equilibrium, matrix $k_\mu^\nu $ is (almost) a constant matrix \cite{onsager}, say $k_{0\mu}^\nu$, and Eq.~(\ref{cd9}) reduces to 
\begin{equation}\label{cd10}
\frac{d\xi_\mu}{dt}=J_\mu\simeq -k_{0\mu}^\nu\xi_\nu
\end{equation}
\noindent By combining now Eq.~(\ref{cd10}) with the first equation of Eqs~(\ref{cd7}), we finally obtain
\begin{equation}\label{cd11}
J_\mu\simeq L_{\mu\nu}X^\nu\qquad{\rm with}\quad
L_{\mu\nu}\equiv k_{0\mu}^\eta{\tilde h}_{0\nu\eta}
\end{equation}
\noindent where ${\tilde h}_{0\mu\nu}$ denotes the inverse of the matrix $ h_0^{\mu\nu}$, i.e., $h_0^{\kappa\mu}{\tilde h}_{0\kappa\nu}=\delta_\nu^\mu$. In literature, matrix $L_{\mu\nu}$ is called {\it the Onsager matrix of the transport coefficients}. The main conclusion of the present analysis is that: {\it the transport relations can be linearized with respect to the thermodynamic forces only when the system is sufficiently close to the (local) equilibrium} (i.e., when the thermodynamic forces are "weak"). {\it In this situation, the distribution function for fluctuations is a Maxwellian. Reversely, the transport relations are necessarily nonlinear in case of non-Maxwellian fluctuation-distribution function}. Let us consider, for example, the case of magnetically confined tokamak-plasmas. After a short transition time, the plasma remains close to (but, it is not in) a state of local equilibrium. Indeed, starting from an arbitrary initial state, the collisions would tend, if they were alone, to bring the system very quickly to a local equilibrium state, described by a Maxwellian distribution function. But slow processes, i.e. free-flow and electromagnetic processes, prevent the plasma from reaching this state. The distribution function for the fluctuations of the thermodynamic quantities also deviates from a Maxwellian preventing the thermodynamic fluxes from being linearly connected with the conjugate forces.

\noindent We would like to end this section by adding some physical remarks. In this paper, we have shown that magnetically confined plasmas {\it tend to relax} towards the mechanical equilibrium following the shortest path, traced out in the space of the parallel generalized frictions. We may ask the following question: "{\it why should the thermodynamic forces, and not other quantities, obey this law ?} ". The answer is, because these quantities are purely {\it non-conservative}. As known, the thermodynamic variables can be classified as {\it conservative} or {\it non-conservative}. A fluctuation of a conservative variable can be "dissipated" only through the boundaries and, due to his severe constraint, its evolutionary-trajectory towards the steady-state may be very complex in the phase-space. On the contrary, a fluctuation of a non-conservative variable can be dissipated freely into the surrounding and its evolutionary-trajectory tends to approximate that of the shortest path. We know that, in thermodynamics, one of the most  dissipative quantity is ${\mathcal P}$, the Glansdorff-Prigigine quantity, given by Eq.~(\ref{c5e})
\begin{equation}\label{e2a}
{\mathcal P}=-{\dot\varrho}\int\Bigl(\frac{d\varsigma}{d\varrho'}\Bigr)^2d\varrho'\le0
\end{equation}
\noindent where, we recall, $\varsigma$ is the arc-parameter in the space of the parallel generalized frictions and $\varrho$ parameterizes  the equation for the shortest path in this space. By noticing that the transport coefficients depend only on the thermodynamic forces $X^\mu$, it turns out that also quantity ${\mathcal P}$ depend only on the parallel generalized frictions. Since the thermodynamic forces are non-conservative variables and quantity ${\mathcal P}$ is defined in the space of the parallel generalized frictions, it is not so surprising to have observed that, in the thermodynamical forces-space, the $X^\mu$ tend to follow the shortest path for reaching the steady-state. This law is so well satisfied experimentally that we can write
\begin{equation}\label{e2b}
J_{\mu}^\alpha=g_{\mu\nu}(X)X^{\alpha\nu}\simeq {\tilde g}_{\mu\nu}(X)X^{\alpha\nu}\qquad [{\rm or,}\quad g_{\mu\nu}(X)\simeq {\tilde g}_{\mu\nu}(X)]
\end{equation}
\noindent during the gradual relaxation of the thermodynamic forces towards the (global) mechanical equilibrium, with 
\begin{equation}\label{e2c}
J_{\mu\ st.state}^\alpha=g_{\mu\nu}(X_{st.state})X_{st.state}^{\alpha\nu}=
{\tilde g}_{\mu\nu}(X_{st.state})X_{st.state}^{\alpha\nu}
\quad [{\rm or,}\ \ g_{\mu\nu}(X_{st.state})= {\tilde g}_{\mu\nu}(X_{st.state})]
\end{equation}
\noindent rigorously valid at the steady-state. We draw the attention to the reader that we prefer to use the expression "the system {\it tends to follow} the shortest path" instead of "the system {\it follows} the shortest path" because, as known, the Universal Criterion of Evolution, expressed by inequality (\ref{e2a}), cannot be derived from a variational principle [Ref. \cite{sonninog}, \cite{prigogine1} and, our results Eqs~(\ref{c6b})]. For this reason, in general, the shortest path is not the trajectory that minimizes the Glansdorff-Prigogine quantity.

\section{Acknowledgments}
\noindent This paper is dedicated to Prof. Pierre Coullet of the Institut Nonlin{\'e}aire de Nice (France). I  joined, in 1992, the research group of Prof. P. Coullet as post doctoral position for working on pattern formation in chemical systems induced by a parametric forcing. It has been a great pleasure and honor to work with him. 

\noindent I would like to pay tribute to the memory of Prof. I. Prigogine who gave me the opportunity to exchange most interesting views in different areas  of thermodynamics of irreversible processes.

\noindent I address a special thanks to Prof. M. Malek Mansour of the Universit{\'e} Libre de Bruxelles (U.L.B.). I have greatly appreciated his beautiful lectures on {\it Dynamics of the Hydrodynamical Fluctuations} held at the U.L.B.


\end{document}